\documentclass[preprint,showpacs,preprintnumbers,amsmath,amssymb]{revtex4}
\usepackage{graphicx}
\usepackage{dcolumn}
\usepackage{bm}
\usepackage{color}

\begin{document}
\title{Statistical models for nucleic acids
}

\author{Marco Zoli}

\address{School of Science and Technology \\  University of Camerino, I-62032 Camerino, Italy \\ marco.zoli@unicam.it}

\begin{abstract}
ds-RNA and standard ds-DNA show specific structural differences in their di-nucleotide steps, piled along the helical axis. Modeling the helices of short fragments by a 3D mesoscopic Hamiltonian model, I use path integral techniques to compute the average helical repeat and show that these structural features are at the origin of the opposite twist-stretch patterns of the A- and B- form.
\end{abstract}

\maketitle

\begin{figure}
\includegraphics[height=7.0cm,width=6.0cm,angle=-90]{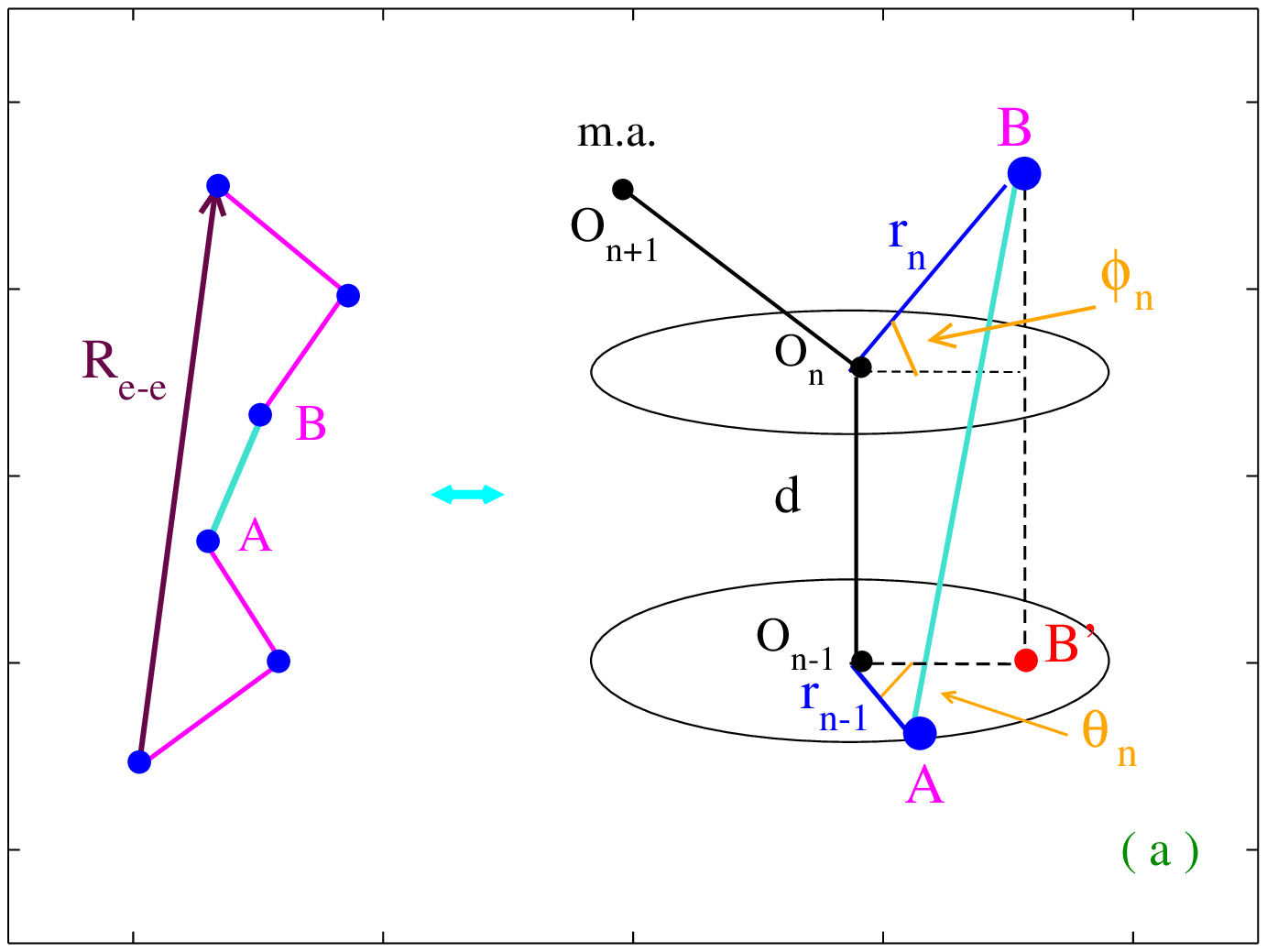}
\includegraphics[height=7.0cm,width=6.0cm,angle=-90]{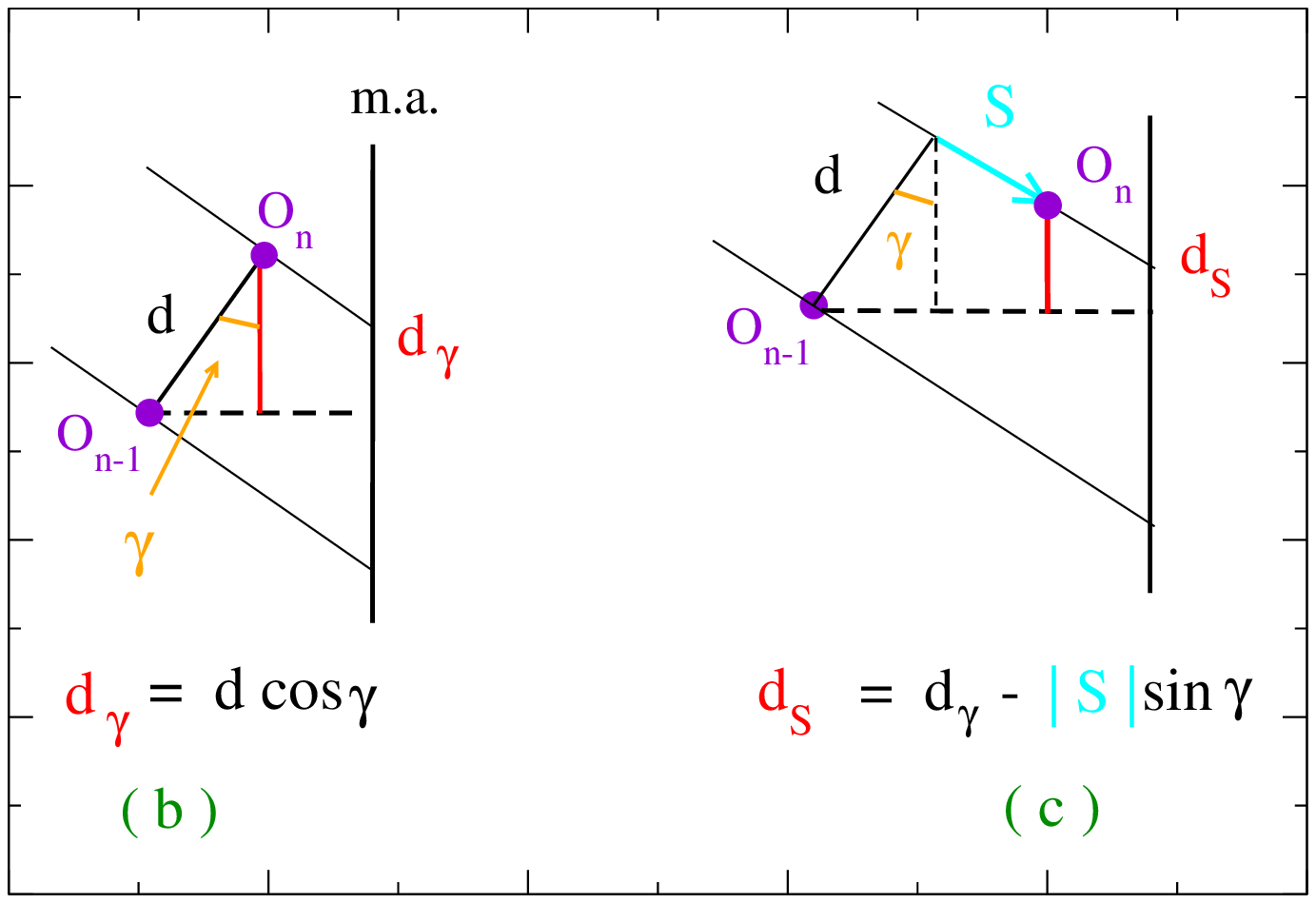}
\caption{\label{fig:1}(Color online)  
(a) 3D model for a helical chain.  $\phi_{n}$ is the bending angle between adjacent $r_{n}$'s in a dimer;  $\theta_n$ is the twist accumulated along the helix.  $\overline{AB}$ is the average dimer distance. 
$R_{e-e}$ is the end-to-end distance. (b) Base pair inclination with respect to the helical axis measured by the  tilt angle $\gamma $: the rise distance (along the molecular axis) $d_{\gamma }$ is shorter than the bare rise distance $d$.
(c) The n-th bp slides on top of the adjacent (n-1)-th bp.  $S$ is taken as negative hence, the rise distance $d_{S}$ gets shorter than $d_{\gamma }$. 
}
\end{figure}

Coarse-grained Hamiltonian models provide useful descriptions of nucleic acid at the level of the base pairs (bps) and permit to calculate the structural properties of helicoidal chains as a function of the environmental conditions. Several studies produced over the last years have investigated denaturation bubbles, thermodynamics, flexibility, force/stretching relations and length distribution functions of linear chains and loops with variable size and sequence \cite{io13b,io14b,io15,io16a,io16b,io17,io18b,io18c,io18a}. These works are based on a 3D model, see schematic in Fig.~\ref{fig:1}(a),  which incorporates the twisting and  bending degrees of freedom between adjacent bps along the molecule stack while the $r_{n}$'s measure the relative distance between complementary bp mates.  
If the average bending angles are close to zero, then
the $n-th$ bp radial fluctuations occur in a plane almost perpendicular to the molecular axis. If this feature is common to a large number of dimers in the chain then there is no overall significant tilt of the bps planes. This picture is appropriate to model the physiological B-form of ds-DNA, in which the helix axis runs through the center of each bp and the bps are stacked per­pendicular to the axis. 
On the other hand, the more compact (and broader) A-form helix adopted by ds-RNA displays two microscopic features, visualized in Fig.~\ref{fig:1}(b), whose entity is dependent on the di-nucleotide step:  i) the bp planes are tilted by the angle $\gamma $ respect to the vertical helical axis and ii) the bps forming a dimer, slide by a distance $S$ past each other. For simplicity it is hereafter assumed that, for a single simulation, $\gamma $ and $S$ are average values distinctive of the chain although local variations are expected in specific sequences.
The occurrence of tilt and slide has the direct consequence to shorten the rise distance along the helical axis, a feature which affects the stretching flexibility.

\section{Method}

The model, for a chain with $N$ bps, is studied by the path integral computational method, discussed in a number of papers, see e.g. refs.\cite{io09,io10,io11,io14a,io14c}. 
In the finite temperature path integration, the $r_n$'s are conceived as trajectories $r_n(\tau)$ where $\tau$  is the Euclidean time varying in the range $[0, \beta]$ and  $\beta$ is the inverse temperature \cite{io97,io03,io05}. Imposing the closure condition $r_n(0)=\,r_n(\beta)$, we can expand $r_n(\tau)$ in Fourier series around  the average helix diameter $R_0$, \, $r_n(\tau)=\, R_0 + \sum_{m=1}^{\infty}\Bigl[(a_m)_n \cos( \frac{2 m \pi}{\beta} \tau ) + (b_m)_n \sin(\frac{2 m \pi}{\beta} \tau ) \Bigr] \,$.
While the Fourier coefficients define in principle all possible choices of fluctuations for any bp, the calculation includes in the statistical partition function a subset of $r_n$'s which fulfill the physical requirements of  the model potential \cite{io11a,io12,io20}.

\section{Results}

Essential to the evaluation of the nucleic acids properties is the evaluation of the integral cutoff $\bar U$ for the Fourier coefficients. To this purpose, let's consider for instance the radial fluctuations marked by the blue dot A in Fig.~\ref{fig:1}(a) and set $j \equiv \, n-1$. Given the Fourier expansion, the initial condition for the average pair mates separation is, $< r_j > =\,R_0$. At any later time $t$, $r_j$ may fluctuate around $R_0$ consistently with the physical properties of the Morse potential (which models the hydrogen bonds between the pair mates) and of the solvent potential (which models the environment surrounding the helical molecule). Accordingly, as the $r_j$ fluctuation is conceived as a constrained Brownian motion \cite{io19,io20a,io21,io22}, we define $P_j(R_0,\, t)$ as the probability  that $r_j$ does not return to $R_0$ until $t$ whereby, for a specific $t$,  $P_j(R_0,\, t)$ is expressed as a sum over the particle histories $r_j(\tau)$ in the time range $[0, t]$:

\begin{eqnarray}
& &P_j(R_0,\, t)=\,   \oint Dr_{1} \exp \bigl[- A_a[r_1] \bigr] 
\cdot \prod_{n=2, \, n\neq j}^{N} \oint Dr_{n}  \exp \bigl[- A_b [r_n, r_{n-1}, \phi_n, \theta_n] \bigr]  \cdot \,  \nonumber 
\\
& & 
\int_{r_j(0)}^{r_j(t)} Dr_{j} \exp \bigl[- A_b[r_j, r_{j-1}, \phi_j,  \theta_j] \bigr]  \cdot
\prod_{\tau=\,0}^{t}\Theta\bigl[r_j(\tau) - R_0\bigr] \, ,
\label{eq:03aa}
\end{eqnarray}

where $A_a[r_1]$ and $A_b[..]$ are the action functionals for the first and $n-th$ bps. The Heaviside function $\Theta[..]$ selects those trajectories $r_j(\tau)$ which stay larger than $R_0$ for any $\tau \in [0, t]$. Hence, the program retains only those sets of Fourier coefficients fulfilling this constraint and adds their contribution to $P_j(R_0,\, t)$. Note that $r_j(\tau)$ describes an open end trajectory ($\tau \leq t < \beta$), while all other $n$ fluctuations are closed along the time axis, $r_n(0) =\, r_n(\beta)$, and therefore their integration measure is $\oint Dr_{n}$. 

\begin{figure}
\includegraphics[height=7.0cm,width=6.0cm,angle=-90]{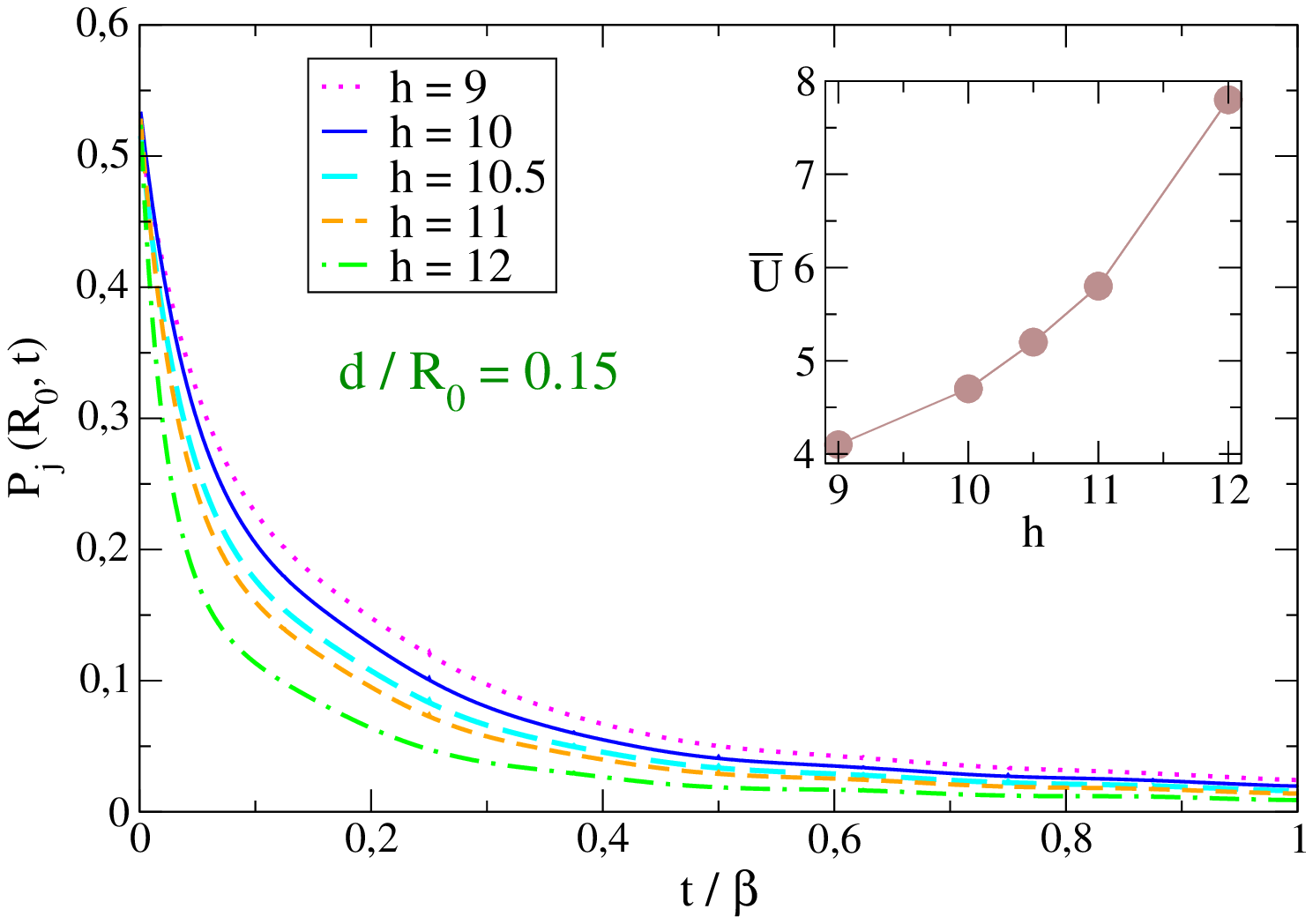}
\caption{\label{fig:2}(Color online)  First-passage probability versus time for the mid-chain bp in equilibrium with the $N-1$ bps at room temperature.  Five twist conformations are considered.  For each $h$, the probability is computed assuming the respective cutoff \={U} (in the inset). 
}
\end{figure}

Let's now define the criterion to calculate $\bar U$. From the path expansion observe that, at $t=\,0$, the $j-th$ trajectory is \, $r_j(0)=\, R_0 + \sum_{m=1}^{\infty}(a_m)_j$. As the Fourier coefficients are integrated on an even domain, the $P_j(R_0,\, 0)$ value obtained from Eq.~(\ref{eq:03aa}) should be $ \sim 1/2$ where the approximation stems from the constraint that large negative coefficients are not allowed by the hard core of the Morse potential. This is the zero time probability to have a radial amplitude larger than $R_0$ and this provides the benchmark to select the meaningful, time dependent probabilities as a function of the cutoff.
Technically, the integration $\int_{r_j(0)}^{r_j(t)} Dr_{j}$  in Eq.~(\ref{eq:03aa}) is carried out by setting a cutoff with tunable $U_j$ and then selecting the value  such that $P_j(R_0,\, 0) \sim 1/2$. As the chain is taken homogeneous, the selected, $U_j \equiv \bar{U}$, holds for all bps in the chain. Generally, for heterogeneous sequences, one can apply the method to derive a set of site dependent cutoffs.
Thus, the maximum amplitude of the bp fluctuations is physically related to the set of model potential parameters and to the specific twist conformation for the molecule. The results for a homogeneous fragment of $21$ bps are displayed  in Fig.~\ref{fig:2}: the time dependent probability is plotted versus time for five average twist conformations defined by $h$ i.e. the number of bps per helix turn. As reported in the inset, the $\bar{U}$ value, which satisfies the zero time probability benchmark, gets larger if the helix unwinds. A physically plausible result.

\begin{figure}
\includegraphics[height=7.0cm,width=6.0cm,angle=-90]{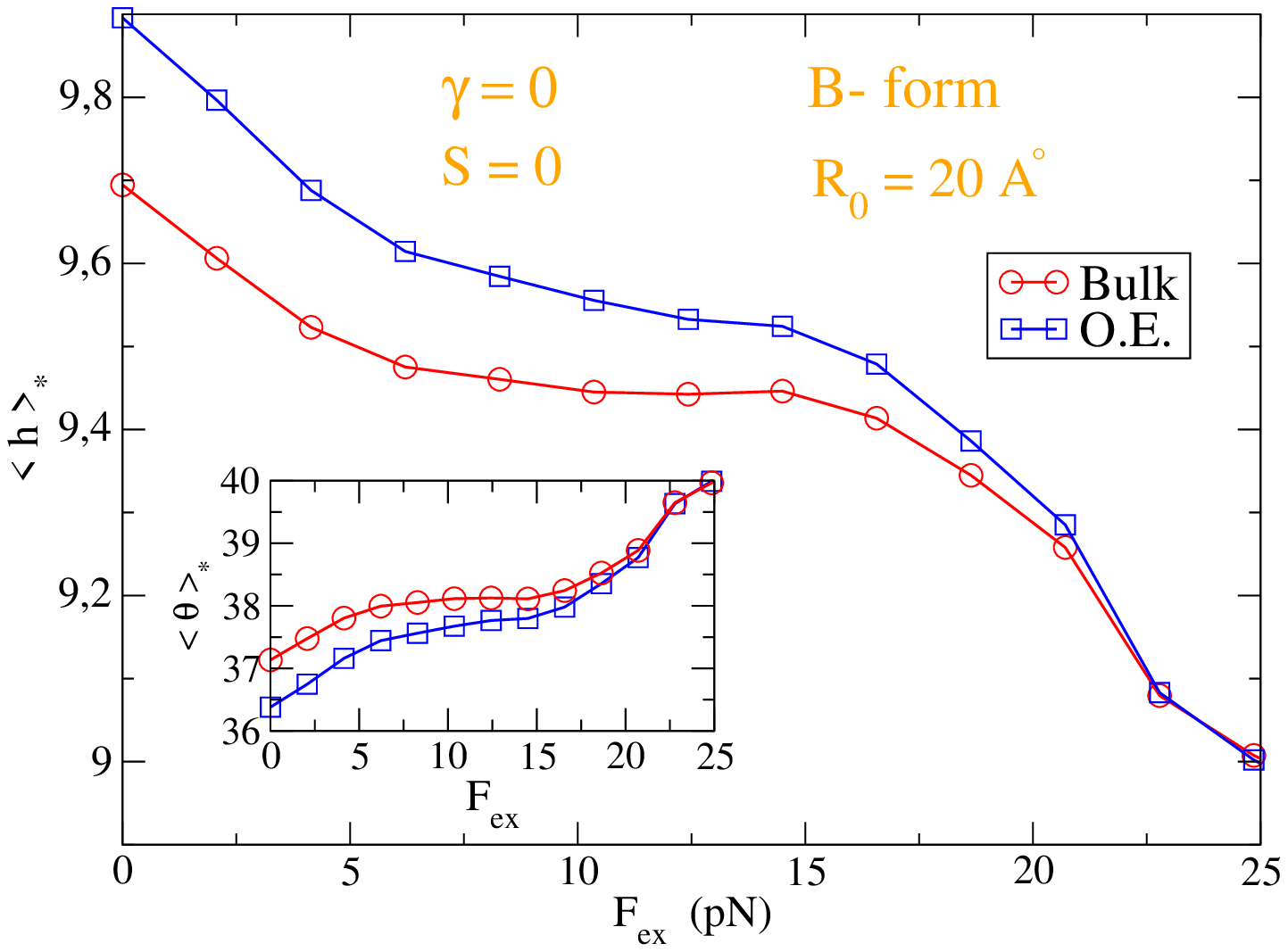}
\includegraphics[height=7.0cm,width=6.0cm,angle=-90]{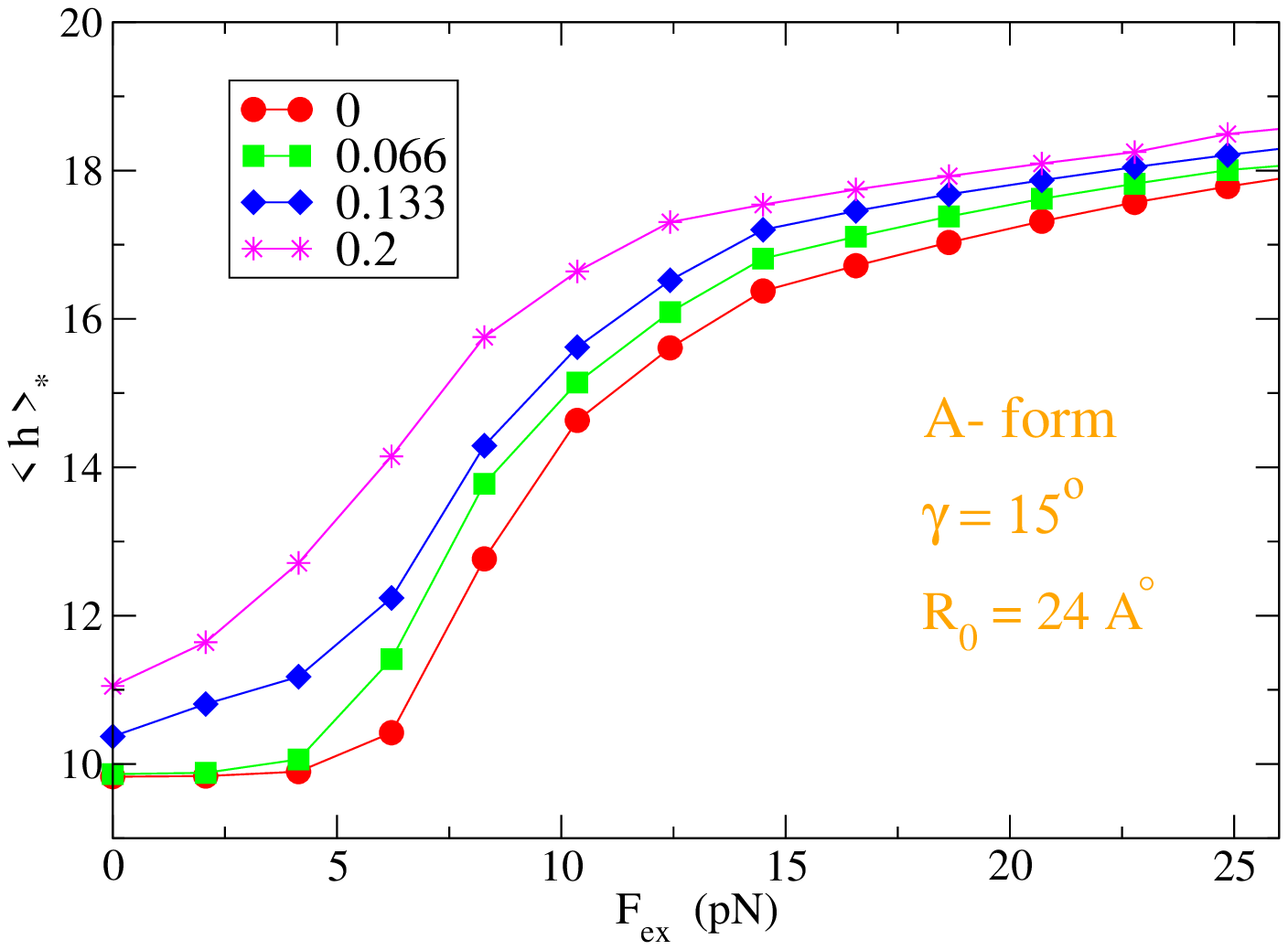}
\caption{\label{fig:3}(Color online)  (a)  Ensemble averaged helical repeat versus external load for the B- form helix. The molecule is considered both with (Open Ends) and without (Bulk) terminal bps.  The inset shows the average twist angle for both cases.  
(b)  Ensemble averaged helical repeat versus external load for the A- form helix.  Four $|S| / d$ ratios are assumed.
}
\end{figure}

After settling the cutoff issue, we focus on the interplay between form and helical conformation of nucleic acids, performing a quantitative analysis of their twist-stretch properties. In particular it is known that the elastic responses of A- and B- DNA structures to external perturbations are strikingly different with A-type and B-type fragments respectively untwisting and over-twisting upon stretching.
To address the twist-stretch relations,  the average helical repeat is computed by performing integrations over the ensemble of bp configurations defined by the partition function for the molecule subjected to a load $F_{ex}$. By minimizing the free energy over a set of possible twist conformations, one selects the equilibrium helical repeat, $< h >_{{*}}$, for a given load. By tuning $F_{ex}$ in the picoNewton range,  the twist profiles are derived for a specific helical molecule. 
The results are shown in Fig.~\ref{fig:3} for the standard B- form of DNA and for the A-form of RNA. 
To emphasize the role of fraying effects, Fig.~\ref{fig:3}(a) displays the ensemble averaged equilibrium helical repeat versus the applied load in two cases: i) the full open ends (O.E.) chain made of $N-1$ dimers and ii) the bulk of the chain made of $N-3$ dimers. In the former case, $< h >_{{*}}$ is significantly larger signaling that the terminal bps strongly affect the overall helix untwisting thus yielding an enhanced flexibility. This holds both in the absence of loads and for moderate loads up to about $20 \,pN$ whereas, for strong $F_{ex}$, the chain end effects tend to vanish as the over-twisting is more pronounced. 
For the A-form, Fig.~\ref{fig:3}(b), a bp inclination $\gamma =\, 15^{o}$ with respect to the helical axis is assumed consistent with X-ray diffraction data and molecular dynamics simulations. Both the zero slide case and three cases with finite $|S| / d$ values are considered.  
All plots show that $< h >_{j^{*}}$ grows under the effect of the stretching load revealing that the bp inclination is the primary cause of the helix untwisting. The effect of the slide is superimposed to that of $\gamma$: by increasing $|S|$, the helix untwisting is larger for all $F_{ex}$ suggesting that the structural features of the dimers determine the overall flexibility of the chain.

\end{document}